\documentstyle[12pt,psfig]{article}
\newcommand{\mbf}[1]{\mbox{\boldmath $ #1 $}}
\newcommand{\q}{\mbf{q}}                             
\newcommand{\dd}{$\Delta$}
\newcommand{\de}{\Delta}
\newcommand{\beq}{\begin{equation}}
\newcommand{\eeq}{\end{equation}}
\newcommand{\beqa}{\begin{eqnarray}}
\newcommand{\eeqa}{\end{eqnarray}}
\newcommand{\beit}{\begin{itemize}}
\newcommand{\eeit}{\end{itemize}}
\newcommand{\benu}{\begin{enumerate}}
\newcommand{\eenu}{\end{enumerate}}
\begin{document}
\begin{titlepage}
 
{\small \sl Physics Letters B}\hfill ECT$^*$-98-003, LBNL-41697R\\[12ex]
 
\begin{center}{\large {\bf
    Transport Simulations with $\pi$ and $\Delta$ 
    In-Medium Properties$^\dagger$}}\\[8ex]
{\sl Johan Helgesson$^{a,\ddagger}$ and J\o rgen Randrup$^b$}\\[2ex]
 
$^a$ECT$^*$, European Centre for Studies in Theoretical Nuclear Physics 
                      and Related Areas, Trento, Italy\\[1ex]
 
$^b$Nuclear Science Division, 
         Lawrence Berkeley National Laboratory,\\
         Berkeley, California 94720, USA\\[6ex]
July 10, 1998  \\[6ex]

{\sl Abstract:}\\
\end{center}
{\small\noindent
Transport simulations including in-medium properties
derived in a microscopic 
$\pi + N N^{-1} + \Delta N^{-1}$ model 
in infinite nuclear matter are presented.
In-medium pion dispersion relations,
partial $\Delta$ decay widths,
pion absorption cross sections
and $\Delta$ cross sections
are incorporated into the transport description
by means of a local-density approximation.
Strong modifications of $\pi$ and $\Delta$
production and absorption rates are found,
but only small effects on pion observables.

\vfill
\\
 
\\
 
\noindent
\footnotesize
$^\dagger$This work was supported 
by the Training and Mobility through Research (TMR) programme 
of the European Community under contract ERBFMBICT950086 
and by the Director, Office of Energy Research,
Office of High Energy and Nuclear Physics,
Nuclear Physics Division of the U.S. Department of Energy
under Contract No.\ DE-AC03-76SF00098.\\
$^\ddagger$Present address: Malm\"{o} University, 
           School of Engineering and Economics,
           205 06 Malm\"{o}, Sweden.
}
\end{titlepage}

\section{Introduction}

In collisions between two heavy nuclei at bombarding energies
from a few hundred MeV up to several GeV per nucleon,
hadronic matter at high density and temperature is formed
and a large number of energetic particles are produced
\cite{Metag,Cassing,Mosel,Ko_review}.
The main features in such collisions 
have been fairly well described by microscopic transport models
employing vacuum properties of resonances and mesons,
such as most versions of BUU and QMD \cite{Cassing,Mosel,Wolf,Aichelin}.
However, $\pi$ mesons, nucleons, and \dd\ isobars 
are strongly interacting particles,
and in infinite nuclear matter
a system of such particles would couple to form spin-isospin modes.
It is quite possible that such in-medium effects 
could play an important role for the transport properties.
First attempts to employ such in-medium modifications 
in transport simulations of nuclear collisions have been carried out.
In Refs.\ \cite{Weise,Bertsch,Giessen,Texas,Tub} 
the employed in-medium properties 
were based on a simple two-level $\de N^{-1}$ model.
A more complete $\pi + N N^{-1} + \Delta N^{-1}$ model 
was used in Refs.\ \cite{sim} 
to derive a consistent set of in-medium quantities
suitable for transport descriptions,
with exploratory transport simulations
reported in Refs.\ \cite{transp_JHJR}.

The previous attempts to include 
$\pi$ and $\Delta$ in-medium properties
have suggested that only small effects 
are found in pion observables.
This is not surprising since most emitted pions
are created at the surface 
where the in-medium effects are small.
However, in-medium modifications of pions and $\Delta$ isobars 
may be important for other observables.
For example, presently large efforts are made
to study effects of possible chiral restoration
at high densities and temperatures,
and in particular by investigating kaon and dilepton spectra
which may carry signatures of in-medium modifications 
of vector meson masses due to chiral restoration 
\cite{Ko_review,KochSong,Rapp,Klingl}.
Since kaons and dileptons are produced 
through multistep processes,
involving pions and $\Delta$'s,
in-medium modifications
of $\pi$ and $\Delta$ quantities 
could implicitly have an important impact.

In this letter we present selected results from transport simulations
that include in-medium quantities obtained from the  
$\pi + N N^{-1} + \Delta N^{-1}$ model 
of Ref.\ \cite{sim}.
The present work constitutes an extension 
of Ref.\ \cite{transp_JHJR},
now including a more complete and consistent set
of in-medium quantities.
Our intention is to
present a qualitative picture 
of some in-medium effects that 
survive the transport dynamics,
while a more complete and systematic presentation
will be given in a subsequent paper.
While the model presented here contains
the main in-medium modifications of pions and $\Delta$ isobars,
a number of additional processes might play a role as well.
For example,
effects originating from a possible partial chiral restoration
have not been included.
Moreover, there might also be in-medium modifications 
of cross sections and decay widths originating
from the fact that the time and volume available 
in collisions and decays are finite.
We expect that the present transport simulations 
contain the most important in-medium features.
Furthermore,
they provide the most consistent tests carried out so far
of in-medium effects within $\Delta N^{-1}$ models.

\section{The Model}
\label{sec_Model}
Our approach is to obtain the medium modifications
by microscopic calculations in uniform matter at various densities
and then incorporate those into a transport treatment
by a local-density approximation \cite{sim,transp_JHJR}.
For simplicity, 
all microscopic quantities are approximated
by their zero-temperature values
and the $\Delta$ width is omitted
in the microscopic treatment of the dispersion relations.
Detailed discussions of these approximations
are given in Refs.\ \cite{sim,transp_JHJR}.


We consider interacting 
nucleons ($N$), delta isobars ($\Delta$), 
pi mesons ($\pi$), and rho mesons ($\rho$)
in a periodic box for $T=0$. 
The in-medium properties are obtained by 
using the Greens function technique,
starting from non-interacting hadrons.
At pion vertices 
we use effective $p$-wave interactions
containing form factors,
$F_{N \pi N}(q)$ and $F_{N \pi \Delta}(q)$.
Since various choices are made in the literature 
we will in this work utilize two different sets.
The first set, denoted FF1, is taken
\beq
       F_{N \pi \alpha}(q) 
    =
       \left[ \frac{2m_\alpha c^2}
                   {m_\alpha c^2 + \sqrt{s}} \right]^{1\over2}
       \frac{\Lambda_\pi^2 - (m_\pi c^2)^2}{\Lambda_\pi^2 - (cq)^2}\ ,
  \qquad
       \alpha = N, \Delta
\eeq
while for the second set, FF2,
the relativistic corrections 
are replaced by an off-shell correction
at vertices including a $\Delta$ isobar 
and neglected at other vertices.
In addition the monopole form 
is approximated by an exponential form.
Thus for FF2 we take
\beq
       F_{N \pi N}(q) 
    =
       {\rm e}^{ - |(cq)^2 - (m_\pi c^2)^2|/ \Lambda_\pi^2 }\ ,
  \qquad
       F_{N \pi \Delta}
    =
       \left[ \frac{ {\q}_0^2 + \kappa^2 }
		   { {\q}_{cm}^2 + \kappa^2} \right]^{1\over2}
       {\rm e}^{ - |(cq)^2 - (m_\pi c^2)^2|/ \Lambda_\pi^2 }\ .
\eeq
For most quantities of interest here the difference between the
monopole and exponential form is practically negligible,
but the exponential form yields better convergence properties 
for calculating the real part of the $\Delta$ self energy.
The relativistic correction 
$2 m_\alpha/[\sqrt{s} + m_\alpha]$
is frequently used in the $\Delta$-hole model 
in connection with pion 
absorption and scattering on nuclei \cite{Oset-Weise,Strottman},
while the form 
$[ \mbf{q}_0^2 + \kappa^2 ]/[ \mbf{q}_{cm}^2 + \kappa^2]$
takes into account the $\Delta$ off-shell correction
and has been used for calculating
vacuum $N+N \rightarrow \Delta + N$ cross sections \cite{ppnD},
and is also frequently included in vacuum $\Delta$ widths
incorporated in transport simulations 
\cite{LiBauer,Kitazoe}.
Since the two form factor choices, FF1 and FF2,
lead to somewhat different in-medium properties,
we have chosen to test both forms.

At the $N \rho N$ and $N \rho \Delta$ vertices 
interactions corresponding to those used at the pion vertices
are used, and at baryon-hole vertices 
effective short-range interactions are included,
moderated by the  correlation parameters 
$g_{NN}'$, $g_{N\Delta}'$, and $g_{\Delta \Delta}'$.

A set of RPA equations for spin-isospin modes,
corresponding to an infinite iteration of
non-interacting pion ($\rho$-meson), nucleon-hole, 
and $\Delta$-hole states,
were derived in Ref.\ \cite{sim}.
From these equations eigenvectors and
eigenenergies are obtained 
for the different spin-isospin modes.
We also calculate total and partial $\Delta$ widths,
$N \pi \rightarrow \Delta$ cross sections,
$NN \leftrightarrow \Delta N$ cross sections
and $\Delta$ spectral functions within the RPA approximation
(explicit expressions are given in Ref.\ \cite{sim}).

\begin{table}[th]
\begin{tabular}{||l|l|l|l|l||}

\hline
\hline

&  FF1 \& FF2
&  FF1 \& FF2
&  \quad FF1
&  \quad FF2
\\
\hline
     $m_N c^2 = 0.940$
&    $f^{\pi}_{NN} = 1.0$
&    $f^{\rho}_{NN} = 6.2$           
&    $g^{\prime}_{NN} = 0.9$
&    $g^{\prime}_{NN} = 0.9$
\\
     $m_{\Delta}c^2 = 1.23$
&    $f^{\pi}_{N \Delta } = 2.2$
&    $f^{\rho}_{N \Delta } = 10.5$
&    $g^{\prime}_{N \Delta} = 0.38$
&    $g^{\prime}_{N \Delta} = 0.43$
\\
     $m_{\pi}c^2 = 0.14$
&    $f^{\pi}_{\Delta \Delta } = 0$
&    $f^{\rho}_{\Delta \Delta } = 0$ 
&    $g^{\prime}_{\Delta \Delta} = 0.50$
&    $g^{\prime}_{\Delta \Delta} = 0.40$
\\
     $m_{\rho}c^2 = 0.77$
&    $\Lambda^{\pi} = 1.0$
&    $\Lambda_{\rho} = 1.5$                            
&    $\Lambda_g = 1.5$
&    $\Lambda_g = 1.5$
\\
\hline
     $m_N^*= m_N$
&    $\rho_0 = 0.153\ \mbox{ fm}^{-3}$    
&    \multicolumn{3}{l||}{ $V_\Delta - V_N \quad
			=\ 0.025 \,	\rho/\rho_0\ ,\quad 
                           \quad \rho \leq \rho_0\ $}
\\

&    $T = 10^{-4}$
&    \multicolumn{3}{l||}{ $V_\Delta - V_N\quad =\ 0.025\ ,
                         \quad\quad\quad\quad \rho > \rho_0\ $}
\\
\hline
\hline
\end{tabular}
\caption{ Parameter values used in microscopic calculations
          (energy units: GeV).
          \protect\label{tab_param} }
\end{table}
The results of the $\Delta N^{-1}$ model 
depend on the model parameters
which cannot be determined uniquely 
by comparison to experimental data. 
In Ref.\ \cite{sim} was chosen a particular set that could reproduce 
vacuum $p p \rightarrow \Delta^{++} n$ cross sections 
and the imaginary part of the empirical $\Delta$-nucleus 
spreading potential of Ref.\ \cite{Hirata}. 
Unfortunately, the established transport treatment 
that we have chosen to compare 
our medium-modified simulations to \cite{LiBauer}
does not contain any effective nucleon mass,
as was utilized in the microscopic treatment 
of Ref. \cite{sim}.
For this reason we have here ignored the effective mass
in the microscopic calculations
(i.e. we have taken $m_N^* = m_N$).
In addition we are using different sets of form factors.
To still reproduce the vacuum 
$p p \rightarrow \Delta^{++} n$ cross section
and the empirical spreading potential of Ref.\ \cite{Hirata}
as well as possible
we have readjusted the $g^\prime$ parameters
(while keeping other parameters as in Ref.\ \cite{sim}).
With the parameter choice in Table \ref{tab_param},
we find that FF1 over predicts Im $V_{\rm spread}$ somewhat,
while FF2 falls below the empirical points.
Furthermore, FF1 yields a slightly larger
$p p \rightarrow \Delta^{++} n$ vacuum cross section
than FF2.
Therefore,
we will often show results obtained with both form factor choices.


The microscopic calculations 
are performed in a system 
where the medium is at rest
and the obtained dispersion relations and decay widths 
refers to this medium frame. 
When incorporating the in-medium quantities 
into the transport formalism, 
this can be effectively taken into account of
by defining a {\em local} medium frame
as the frame 
where the local flow velocity vanishes.
The pionic Hamiltonians\footnote{   
   In this work we explicitly propagate 
   both the lower and the upper pionic modes,
   as was also done in Ref.\ \cite{Texas}.
   This differs from Refs.\ \cite{Giessen,Tub}
   where only a single average pionic mode was incorporated.
   The importance of treating 
   the two pionic modes equally
   was discussed in Refs.\ \cite{sim,PRCcom} 
.} 
are deduced from the density-dependent dispersion relations 
in infinite nuclear matter 
by a local-density approximation
and are parameterized 
and utilized in the local medium frame. 

\begin{figure}[th]  
\centerline{\psfig{file=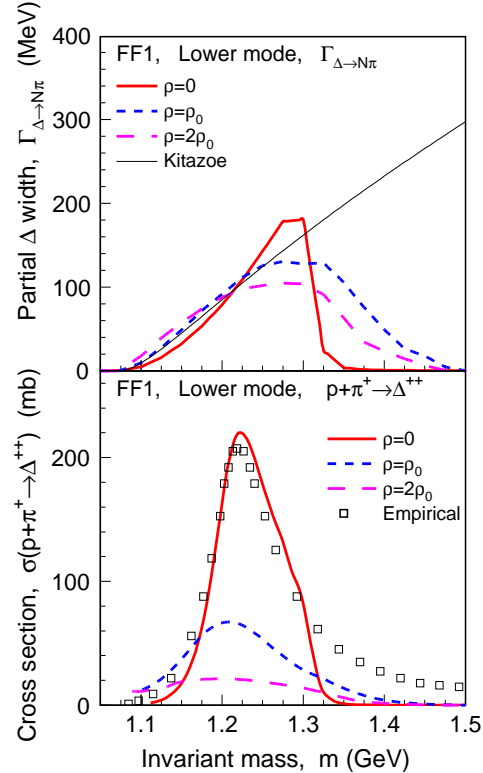,height=11.0cm,angle=0}}
\caption{ Partial $\Delta$ decay widths 
          for the lower pionic mode $\tilde{\pi}_1$ (upper panel) 
          for zero $\Delta$ momentum, and
          pion absorption cross sections
          for the process 
          $p \tilde{\pi}_1^+ \rightarrow \Delta^{++}$
          (lower panel).
          The cross sections are obtained for the special case when the
          $p \tilde{\pi}_1^+$ c.m.\ frame is equal to the rest frame
          of the medium.
          The curve labeled Kitazoe represents the parameterization
          in Ref.\ \protect\cite{Kitazoe},
          and the empirical cross section for $\pi^+$p scattering
          are estimated from Fig.\ 2.2 
          in Ref.\ \protect\cite{EricsonWeise}.
         }
\label{fig_pGam-XSNpi}
\end{figure}

The pionic modes are created in the $\Delta$ decays 
which are governed by partial $\Delta$ widths.
As illustrated in Fig.\ \ref{fig_pGam-XSNpi},
the in-medium widths 
are different than the free width.
This is mainly because of different phase space
and because the pion component on the pionic modes 
no longer is unity;
there are also $\Delta N^{-1}$ and $N N^{-1}$  components.
Note particularly that the width for the decay   
     to the lower mode decreases and approaches zero 
     for invariant masses above $m \sim 1.3$ GeV/$c^2$.
     This is because the pion component vanishes 
     and the collectivity disappears 
     when the mode enters the band 
     of non-collective $\Delta N^{-1}$ modes.
     This differs from the findings in Ref.\ \cite{Texas}
     which did not include a $\Delta N^{-1}$ continuum.
     Otherwise Ref.\ \cite{Texas} found qualitatively similar
     partial widths, with quantitative differences depending
     on the specific form factors and parameter sets.
Another feature of the $\Delta$ width in the medium  
is that it depends explicitly 
on the $\Delta$ momentum in the medium,
in addition to the $\Delta$ energy.
This is especially important for the partial widths.
To this end we calculate and store in a large table 
the partial $\Delta$ decay widths
for discrete values 
of invariant mass, momentum, density and angle.
The $\Delta$ then decays in the medium frame 
according to linear interpolation in the table of decay widths
(for details see Ref.\ \cite{transp_JHJR}).

The pionic modes may be reabsorbed through the process
$ \tilde{\pi}  N \rightarrow \Delta$.
In the standard transport treatment,
the vacuum reabsorption cross section
may be written as an interaction factor 
times a Breit-Wigner factor containing the free $\Delta$ width. 
The in-medium pion absorption cross section
has a similar structure.
For the transport treatment we make use of the fact that
the interaction factor
is also used for calculating the partial $\Delta$ width,
and we use the full in-medium $\Delta$ width in the
Breit-Wigner factor.
Examples of $N  \pi \rightarrow \Delta$
cross sections for FF1 are presented in Fig.\ \ref{fig_pGam-XSNpi}.

The transport simulations are based 
on the hadronic transport model of Li and Bauer \cite{LiBauer}.
In this model the dynamical evolution 
of a heavy-ion collision 
is described  by a set of coupled transport equations, 
explicitly treating nucleons, $\Delta$ isobars and pions.
In this treatment
the inelastic $N N$ cross section is parameterized following
Ref.\ \cite{VerWest}.
In the medium-modified simulations 
we will replace the VerWest and Arndt vacuum cross sections 
for the processes
$N N \leftrightarrow \Delta N$
by the density dependent cross sections 
obtained from the microscopic calculations.
These calculated cross sections constitute in vacuum 
a somewhat cruder approximation 
than the fit obtained in Ref.\ \cite{VerWest} 
where three independent functions $\sigma_{II'}$ were used.
Since our aim in this work is to test in-medium effects,
we will compare to transport simulations 
based on the model described in \cite{LiBauer},
but with the VerWest and Arndt 
$N N \leftrightarrow \Delta N$ cross sections 
replaced by our calculated cross sections at zero density. 
This we will refer to as a {\em standard simulation}.
The microscopic cross sections 
are calculated for the case 
when the c.m.\ system of the colliding particles
coincides with the rest system of the medium.
To keep the transport treatment as simple as possible 
these cross sections are then used 
also when this c.m.\ system differs from the local medium frame.

One important medium modification of the
$N N \rightarrow \Delta N$ cross section
originates from the $\Delta$ spectral function
which is modified in the medium.
This spectral function contains 
the full in-medium $\Delta$ width,
which is calculated self consistently.
The $\Delta$ spectral function 
is also used in the transport treatment 
to determine the invariant $\Delta$ mass,
once a $N N \rightarrow \Delta N$ collision has occured.
The total in-medium $\Delta$ width 
leads at high densities to
a substantial reduction of the
$N N \rightarrow \Delta N$ cross section
for moderate and large c.m.\ energies
and to an enhancement for low c.m.\ energies.
This differs from the results found in Refs.\ \cite{Bertsch,Wu}
where an enhancement was found for large densities.  
However, as noted in Ref.\ \cite{Bertsch},
the result is very
sensitive to the $g^\prime$ parameter values.
In Ref.\ \cite{Wu} it was shown that the enhancement found in
Ref.\ \cite{Bertsch} is weakend by the inclusion of the full $\Delta$
width in the pion polarization function.
Since our calculations employ large values of the
$g^\prime$ parameters\footnote{
     Since the $g^\prime$ parameters are multiplied by
     form factors, there is not a one-to-one correspondance 
     between the values used in Refs.\ \protect\cite{Bertsch,Wu} and
     those used here.}
and in several aspects
differ from Refs.\ \cite{Bertsch,Wu},
we find a net surpression of $\sigma(NN \rightarrow \Delta N)$.
Most importanly,
   we include selfconsistently the full in-medium $\Delta$
   width (including the $\Delta \rightarrow N + NN^{-1}$ contribution 
   which is important at low $\Delta$ mass and high densities)
  and we perform an integration over $\Delta$ masses including
   the in-medium $\Delta$ spectral function.

The cross section for the process 
$N \Delta \rightarrow N N$
is obtained and implemented analogously to the
$N N \rightarrow \Delta N$ cross section.
Note though that the absence 
of the $\Delta$ spectral function
simplifies the treatment.
However, another complication arises 
because the cross section also depends 
on the invariant mass of the colliding $\Delta$, 
in addition to the c.m.\ energy.
We have empirically found that
$\sigma(\sqrt{s};m) \approx \sigma(\sqrt{s'};m_\Delta)$,
with
$\sqrt{s'} = \sqrt{s} - 0.778 (m - m_\Delta)$
yields a good approximation.
The $N \Delta \rightarrow N N$ cross sections show only 
a weak dependence on the nuclear density.

\section{Results}
\label{sec_Res}
In this section we present results from BUU simulations
containing the in-medium quantities as discussed 
in Sect.\ \ref{sec_Model}.
The purpose of these simulations is to elucidate 
the effects of the included in-medium properties.
Therefore, to make the treatment simple
and the signals as clean as possible,
we have ignored some processes 
that may be of equal importance when comparing 
to experiment.
We have performed simulations for central
$^{197}$Au + $^{197}$Au
collisions at 1.0 $A$ GeV
and we have obtained results from standard simulations 
and from three different medium-modified simulations, 
denoted M1--M3.
M1 contains in-medium modifications 
of only pion dispersion relations 
and partial $\Delta$ decay widths.
M2 contains in addition to M1 also the in-medium
pion reabsorption cross section 
$\sigma(N \tilde{\pi} \rightarrow \Delta)$.
Finally, in M3 also the in-medium 
$NN \leftrightarrow \Delta N$ 
cross sections are added. 
All simulations were performed 
with 250 test particles per nucleon
utilizing a mean field
$U(\rho) = -0.218(\rho/\rho_0) + 0.164 ((\rho/\rho_0)^{4/3}$ GeV.

\begin{figure}[th]  
\centerline{\psfig{file=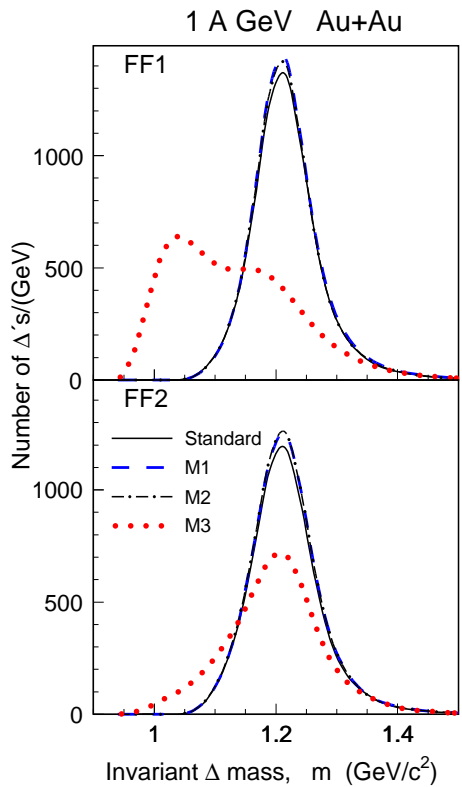,height=11.0cm,angle=0}}
\caption{ Distribution of invariant mass 
          for $\Delta$'s produced in $NN$ collisions.
         }
\label{fig_InvDmDistr-l2}
\end{figure}    
The production rate of $\Delta$ isobars 
from $NN$ collisions  depends on the 
$NN \leftrightarrow \Delta N$ cross sections\footnote{
     The standard simulations 
     with FF1 and FF2 
     yield slightly different results 
     due to slightly different input vacuum 
     $NN \leftrightarrow \Delta N$ cross sections. }
and the density and energy distributions of nucleons.
When the density dependent in-medium
$NN \rightarrow \Delta N$ cross sections are incorporated
(simulation M3) the $NN \rightarrow \Delta N$ rates
are reduced, due to a smaller in-medium cross section
at high densities.
For FF1 this effect is small,
due to a strong enhancement of the cross section at low 
$\sqrt{s}$-energies,
while the reduction for FF2 is substantial.

The in-medium $N N \rightarrow \Delta N$ cross section,
together with the $\Delta$ spectral function
also lead to strong modifications 
of the invariant $\Delta$ mass distributions.
Figure \ref{fig_InvDmDistr-l2} shows that for simulation M3, 
there is a strong reduction of $\Delta$ isobars 
with invariant mass around $m_\Delta=1232$ MeV,
and an enhancement at lower masses. 
Note particularly that the invariant mass distribution 
is non-vanishing also for masses below $m=m_N+m_\pi$.
These $\Delta$'s cannot decay to a pionic mode 
and the only reabsorption channel is 
$N\Delta~\rightarrow~NN$. 
As a result, the net number of pions is somewhat reduced.
The effects found with FF1 are similar
to  FF2
but show more pronounced and peculiar enhancement 
of low mass $\Delta$'s.
This originates from the microscopic calculations
which for FF1 show strong effects of pion condensation
for densities around twice normal nuclear density
(while FF2 does not lead to pion condensation 
 at the densities probed in the applied transport simulations).
In the microscopic calculations,
this phenomenon manifest itself as large enhancements 
of the $\Delta$ decay width to low-energy nucleon-hole modes
which in turn leads to a strong enhancement of the $\Delta$ spectral
function around $m \approx 1$ GeV and corresponding effects
in the $N N \rightarrow \Delta N$ cross section.
These effects survive the transport treatment and 
we find strong enhancements of the number of produced $\Delta$'s
in the invariant mass region around 1 GeV.
The net effect on the number of produced pions is a slight reduction.

\begin{figure}[th]  
\centerline{\psfig{file=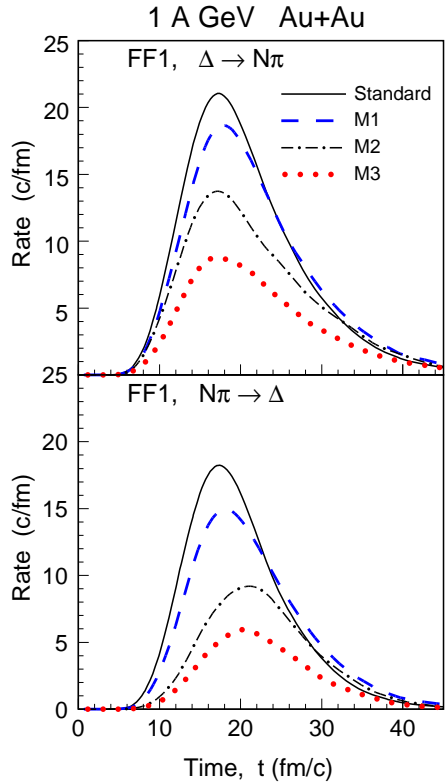,height=11.0cm,angle=0}}
\caption{ The time dependence of the rates for
          $\Delta$ decay (upper panel), and
          the time dependence of the rates for
          pion reabsorption (lower panel).
         }
\label{fig_Rate-l2}
\end{figure}

Inspecting the rate of produced pions from $\Delta$ decay
(upper panel of Fig.\ \ref{fig_Rate-l2}),
we find that for FF1 there is in simulation M1
a slight reduction as compared to the standard treatment,
due to a reduction in the partial $\Delta$ width 
at high densities\footnote{
   For the medium-modified simulations  
   we find that about 95\% 
   of the emitted pions originate from the lower pionic mode. }.
FF2, which for $m \geq m_\Delta$ has only about 60\% 
of the corresponding width obtained for FF1,
gives a strong reduction in the pion production rate.
The further reduction in the $\Delta$ decay rate 
for simulations M2 to M3 reflects 
that there are fewer produced $\Delta$'s in the system.
The simulation M2 illustrates this effect 
due to a reduction of the process 
$N \tilde{\pi} \rightarrow \Delta$ 
while for simulation M3 the reduction is due 
to fewer produced  $\Delta$'s, 
in the mass region $m>m_N+m_\pi$, 
from $NN$ collisions.

The pion reabsorption rate
(lower panel of Fig.\ \ref{fig_Rate-l2})
depends on the density of pions and nucleons and on the
reabsorption cross section.
The reduction in the rate for simulation M1 
is due to the fact that fewer pions are produced from $\Delta$ decay.
There is a further strong reduction 
when including also the in-medium 
$N  \tilde{\pi} \rightarrow \Delta$ cross section (M2). 
This is because this cross section 
is suppressed at high densities
(see Fig.\ \ref{fig_pGam-XSNpi}).
The pion reabsorption rate is reduced 
even further for simulation M3.
This is a secondary effect 
of the fewer produced $\Delta$'s
which then yield fewer pions.

For FF2 the net number 
of $\Delta$ isobars and pions present in the system 
at different times
varies much less with the in-medium input 
than the production and absorption rates.
This is because different effects balance each other
and because the production and absorption
depends on particle densities
which yields an ``automatic'' regulation of the net number.
Note particularly that the enhancement
of pions due to the low reabsorption cross section
at high densities is compensated by a low production
of $\Delta$ isobars by a likewise 
low $NN \rightarrow \Delta N$ cross section at high densities.
This compensation is not accidental
as it depends mainly on the
$\Delta$ in-medium spectral function 
which is included in both cross sections.
The net effect is a reduction of the number of emitted pions
of only about 3-10\%
for simulation M3 as compared to the result 
from the standard simulation.
For FF1 the number of present $\Delta$'s 
varies somewhat more with the different types
of medium-modified simulations.
These variations originates 
from the microscopic onset of pion condensation,
as briefly discussed above.

\begin{figure}[th]  
\centerline{\psfig{file=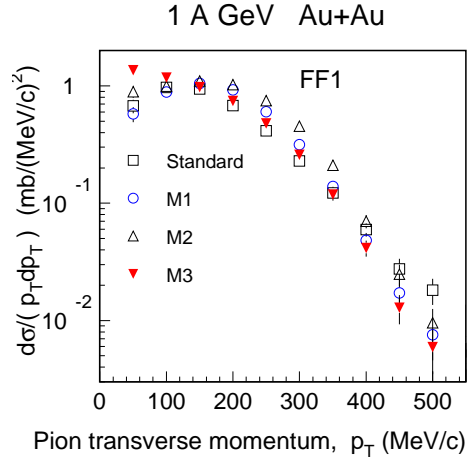,height=6.5cm,angle=0}}
\caption{ The transverse momentum spectrum for neutral pions
          in the central rapidity interval $-0.20 < y_{\rm cm} < 0.20$. 
          The error bars represent the statistical errors.
          The results obtained for FF2 are very similar. 
         }
\label{fig_pipT-l2}
\end{figure}

In-medium effects show up also in the pion energy spectrum,
although to a rather small degree.
Transverse momentum spectra
$d\sigma/{p_T}{dp_T}$
have been obtained at impact parameter $b=1.0$~fm 
assuming that all collisions with an impact parameter up to 
$b_{\rm max}$=$2 \, r_0 A^{1/3}_{\rm Au}$ contribute equally. 
The effect seen in the simulations 
incorporating the in-medium properties, 
as compared to the standard simulations
(see Fig.\ \ref{fig_pipT-l2}),
is a modest but significant enhancement 
at low transverse momenta ($p_T < 250$ MeV/$c$)
and a reduction at higher momenta ($p_T>$350 MeV/$c$),
corresponding to a reduction of the effective transverse temperature.
We find the enhancement of low-energy pions encouraging, 
even though this effect cannot alone explain the enhancement
seen in experimental spectra (Fig.\ 1 of Ref.\ \cite{taps}).
It should be emphasized that
the obtained spectra are not entirely suitable 
for quantitative comparison with experimental spectra
partly because of the simple impact-parameter averaging employed,
and partly because the incorporation of higher nucleon resonances 
and additional medium effects might be of importance. 
Note, for example, that higher-lying resonances 
contribute to both low and high energy pions 
through the two-pion and one-pion emission channels, 
respectively \cite{Giess-2pi_ch}.

\section{Summary}
In this letter, we have presented selected results
from transport simulations incorporating in-medium properties
of pions and $\Delta$ isobars.
In particular, we have included in-medium pion dispersion relations,
partial $\Delta$ widths, pion reabsorption cross sections,
$NN \leftrightarrow \Delta N$ cross sections 
and $\Delta$ spectral functions.
These in-medium quantities have been calculated microscopically
from the $\Delta N^{-1}$ model presented in Ref.\ \cite{sim}
and incorporated into the transport formalism 
of Li and Bauer \cite{LiBauer} by a local-density approximation.
The medium-modified simulations presented in this work 
show strong effects on properties not directly observable 
during the collision process,
such as pion and $\Delta$ production and reabsorption rates,
but only minor effects on spectra of emitted pions.
This is rather reasonable since most of the emitted pions 
are produced at the surface at low densities 
where the in-medium effects are quite small.
However, in an energetic nucleus-nucleus collision 
also other particles, not studied in this work, 
are produced in multistep processes,
where $\Delta$'s and pions act as intermediate particles.
Thus, modified properties of $\Delta$'s and pions 
in the nuclear medium 
might be important to consider 
when studying production of such secondary particles.




\small
\begin{thebibliography}{99}
\bibitem{Metag} V.\ Metag, Nucl.\ Phys.\ {\bf A533} (1993) 283c.

\bibitem{Cassing} W.\ Cassing, V.\ Metag, U.\ Mosel, and K.\ Niita, 
		  Phys.\ Rep.\ {\bf 188} (1990) 363.

\bibitem{Mosel} U.\ Mosel, Ann.\ Rev.\ Nucl.\ Part.\ Sci.\ {\bf 41} (1991) 29.
  
\bibitem{Ko_review} C.M.\ Ko and G.Q.\ Li, 
                    J.\ Phys.\ {\bf G22} (1996) 1673;
                    C.M.\ Ko, V.\ Koch, and G.\ Li, 
                    Ann.\ Rev.\ Nucl.\ Part.\ Sci.\ {\bf 47} (1997) 505.

\bibitem{Wolf} Gy.\ Wolf, G.\ Batko, W.\ Cassing, U.\ Mosel, K.\ Niita,
	and M.\ Sch\"afer, Nucl.\ Phys.\ {\bf A517} (1990) 615.

\bibitem{Aichelin} J.\ Aichelin,  Phys.\ Rep.\ {\bf 202} (1991) 233.

\bibitem{Weise} G.E.\ Brown, E.\ Oset, M.\ Vincente Vacas, and W.\ Weise, 
		Nucl.\ Phys.\ {\bf A505} (1989) 823.

\bibitem{Bertsch} G.F.\ Bertsch, G.E.\ Brown, V.\ Koch, and B.-A.\ Li,
		Nucl.\ Phys.\ {\bf A490} (1988) 745.

\bibitem{Giessen} W.\ Ehehalt, W.\ Cassing, A.\ Engel, U.\ Mosel, 
                  and Gy.\ Wolf,
		  Phys.\ Lett.\ {\bf 298B} (1993) 31.

\bibitem{Texas} L.\ Xiong, C.M.\ Ko, and V.\ Koch,  
		Phys.\ Rev.\ {\bf C47} (1993) 788.

\bibitem{Tub} C.\ Fuchs, L.\ Sehn, E.\ Lehmann, J.\ Zipprich,
	      and A.\ Faessler,
	      Phys.\ Rev.\ {\bf C55} (1997) 411. 	

\bibitem{sim} J.\ Helgesson and J.\ Randrup,
		 Ann.\ Phys.\ (N.Y.) {\bf 244} (1995) 12;
		 Nucl.\ Phys.\ {\bf A597} (1996) 672.

\bibitem{transp_JHJR} J.\ Helgesson and J.\ Randrup,
	              Phys.\ Lett.\ {\bf B411} (1997) 1;
                      J.\ Helgesson,    
                      Proc.\ 36th Int.\ Winter Meet.\ on Nucl.\ Phys., 
                      Bormio, Italy, Ric.\ Sci.\ Educ.\ 
                      Permanente Suppl.\ {\bf 112} (1998) 111.

\bibitem{KochSong} V.\ Koch and C.\ Song,
                   Phys.\ Rev.\ {\bf C54} (1996) 1903.

\bibitem{Rapp} R.\ Rapp, G.\ Chanfray, and J.\ Wambach,
               Nucl.\ Phys.\ {\bf A617} (1997) 472.

\bibitem{Klingl} F.\ Klingl, N.\ Kaiser, and W.\ Weise,
               Nucl.\ Phys.\ {\bf A624} (1997) 527.

\bibitem{LiBauer} B.-A.\ Li and W.\ Bauer,
		  Phys.\ Rev.\ {\bf C44} (1991) 450.

\bibitem{Oset-Weise} E.\ Oset and W.\ Weise,
        Phys.\ Lett.\ {\bf 77B} (1978) 159;
        Nucl.\ Phys.\ {\bf A319} (1979) 477;
        Nucl.\ Phys.\ {\bf A329} (1979) 365.
 
\bibitem{Strottman} E.\ Oset, L.L.\ Salcedo, and D.\ Strottman,
        Phys.\ Lett.\ {\bf 165B} (1985) 13;
        C.\ Garcia-Recio, E.\ Oset, L.L.\ Salcedo, D.\ Strottman, 
        and M.J.\ Lopez,
        Nucl.\ Phys.\ {\bf A526} (1991) 685.
 
\bibitem{ppnD}  V.F.\ Dmitriev, O.\ Sushkov, and C.\ Gaarde, 
                 Nucl.\ Phys.\ {\bf A459} (1986) 503.
 
\bibitem{Kitazoe} Y.\ Kitazoe, M.\ Sano, H.\ Toki, and S.\ Nagamiya,
		 Phys.\ Lett.\ {\bf B166} (1986) 35.

\bibitem{Hirata} M.\ Hirata, J.H.\ Koch, F.\ Lenz, and E.J.\ Moniz,
		 Ann.\ Phys.\ (N.Y.) {\bf 120} (1979) 205.

\bibitem{PRCcom}       J.\ Helgesson and J.\ Randrup,
                       Phys.\ Rev.\ {\bf C56} (1997) 1187.

\bibitem{EricsonWeise} T.\ Ericson and W.\ Weise, Pions and Nuclei 
                       (Clarendon Press, Oxford, 1988).

\bibitem{VerWest} B.J.\ VerWest and R.A.\ Arndt
		  Phys.\ Rev.\ {\bf C25} (1982) 1979.

\bibitem{Wu} J.Q. Wu and C.M.\ Ko,  Nucl.\ Phys.\ {\bf A499} (1989) 810.

\bibitem{taps} O.\ Schwalb {\em et al.}, Phys.\ Lett.\ {\bf B321} (1994) 20.
 
\bibitem{Giess-2pi_ch} S.\ Teis, W.\ Cassing, M.\ Effenberger, 
                       A.\ Hombach, U.\ Mosel, and  Gy.\ Wolf,
                       Z.\ Phys.\ {\bf A356} (1997) 421.

\end{thebibliography}
\end{document}